\documentclass[review]{elsarticle}
\usepackage{lineno,hyperref}
\usepackage{amsmath,amssymb,amsfonts} % Typical maths resource packages
\usepackage{float}
\modulolinenumbers[5]

\journal{Review of Scientific Instruments}
\begin{document}

\begin{frontmatter}
\title{Comments on the radial distribution of charged particles in a magnetic field}

\author{H. Backe \fnref{cor}}
\author{}
\fntext[cor]{corresponding author; E-Mail: backe@kph.uni-mainz.de}
\address{Johannes Gutenberg-University Mainz, Institute for Nuclear Physics, \\D-55099 Mainz, Germany}

\begin{abstract}
Magnetic guiding fields in combination with energy dispersive
semiconductor detectors have been employed already more than 50
years ago for in-beam internal conversion electron spectroscopy.
Even then it was recognized that efficiency modulations may appear
as function of the electron energy, arising when electrons hit a
baffle or miss the sensitive area of the detector. Current high
precision beta decay experiments of polarized neutrons with
conceptional similar experimental devices resulted in a detailed
study of the point spread function (PSF). The latter describes the
radial probability distribution of mono-energetic electrons at the
detector plane. Singularities occur as function of the radial
detector coordinate which have been investigated and discussed by
Sjue at al. (Rev. Scient. Instr. 86, 023102 (2015)), and Dubbers
(arXiv:1501.05131v1 [physics.ins-det]). In this comment a rather
precise numerical representation of the PSF is presented and
compared with approximations in the mentioned papers.
\end{abstract}


\end{frontmatter}

\linenumbers

\section{Introduction}
In two recent papers \cite{SjuB15, Dub15} the radial spread of
charged particles moving in a solenoidal magnetic guiding field
has been investigated. The physical background behind this attempt
is based on the fact that the distribution of the particles at a
detector with finite radius is a potential source of systematic
errors in high precision experiments. To the latter belongs the
measurement of the beta asymmetry in the decay of polarized
neutrons, see, e.g., Ref. \cite{DubR14}. In principle, this fact
is known since long. In the late 60th of the last century nuclear
spectroscopists developed solenoidal transport systems, equipped
with Si(Li) detectors as energy dispersive elements, for in-beam
internal conversion electron spectroscopy. It was pointed out
already in one of the first publications in detail \cite{KlaR72}
that the phase relation between polar emission angle and the
radial coordinate at a circular Si(Li) detector, or a baffle
between target and detector, results in unwanted fluctuations of
the transmission probability as function of the electron energy.
Efficiency modulations on the 10 \% level were reported, e.g., in
Ref. \cite{KotB70}. In order to overcome this problem, the
unwanted wiggles were averaged out by wobbling the magnetic field
strength, see, e.g., also Ref. \cite{LinL75, BacR78}.

From the work of Sjue et al. \cite{SjuB15} and Dubbers
\cite{Dub15} it is now well known that such efficiency modulations
originate from singularities which appear in the so-called
mono-energetic point spread function (PSF) as function of the
radius coordinate. These singularities have been mathematically
treated in the mentioned papers by means of different approaches.
The probability density at the detector plane is presented by Sjue
et al. with the aid of an integral equation while explicitly by
Dubbers, cf. Eq. (14) \cite{SjuB15} and Eq. (12) \cite{Dub15},
respectively. The ways to find solutions are quite different. Sjue
et al. solve the integral equation numerically, while Dubbers
presents and discusses analytical approximations.

This contribution describes an alternative approach with which
numerical solutions of arbitrary accuracy for the PSF can be
obtained. It is based on the mathematically correct parameter
representation of both, the radius coordinate $R(\cos\theta)$ and
the probability density $dP(\cos\theta)/dR$ at the detector plane.
Parameter is $\cos\theta$ which intrinsically is a function of the
polar emission angle $\theta$ at the source. Notice that for
rotational symmetry $d(\cos\theta)$ is just proportional to the
solid angle element $d\Omega$. In the next section
\ref{MathBackbround} first some mathematical details are
described. Results will be compared in section
\ref{ResultsDiscussion} with those presented in Ref. \cite{SjuB15,
Dub15}. The paper closes with conclusions in section
\ref{Conclusions}.

%%%%%%%%%%%%%%%%%%%%%%%%%%%%%%%%%%%%%%%%%%%%%%%%%%%%%%%%%%%%%%%%
\begin{figure}[tb]
%\hspace{3.0 cm}
\centering
    \includegraphics[angle=0,scale=0.65,clip] {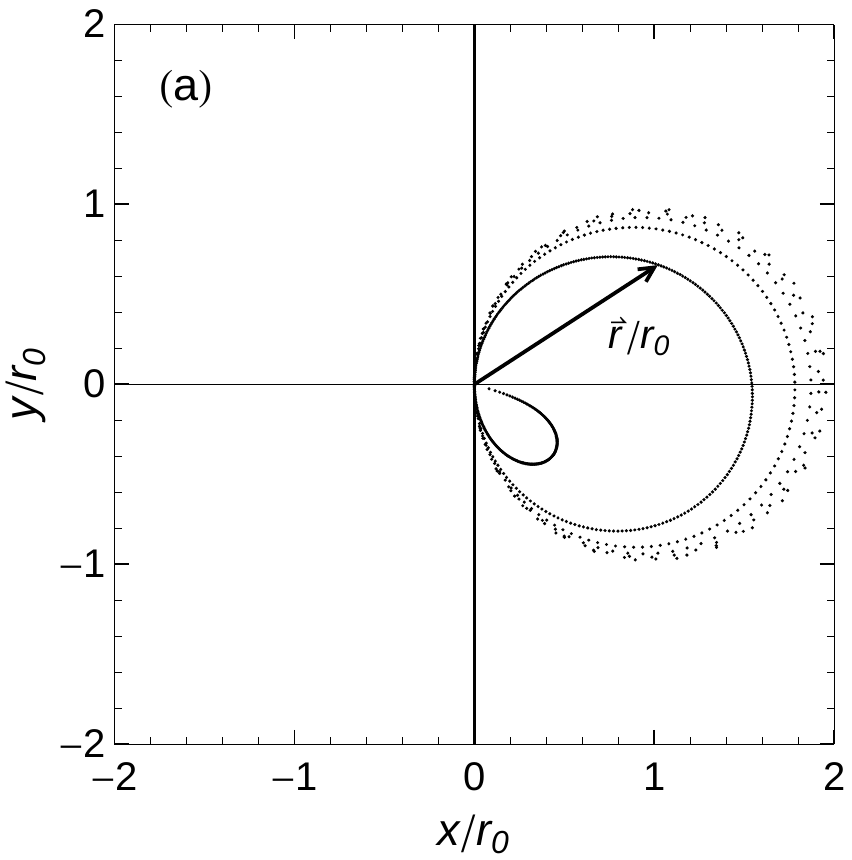}
    \includegraphics[angle=0,scale=0.65,clip] {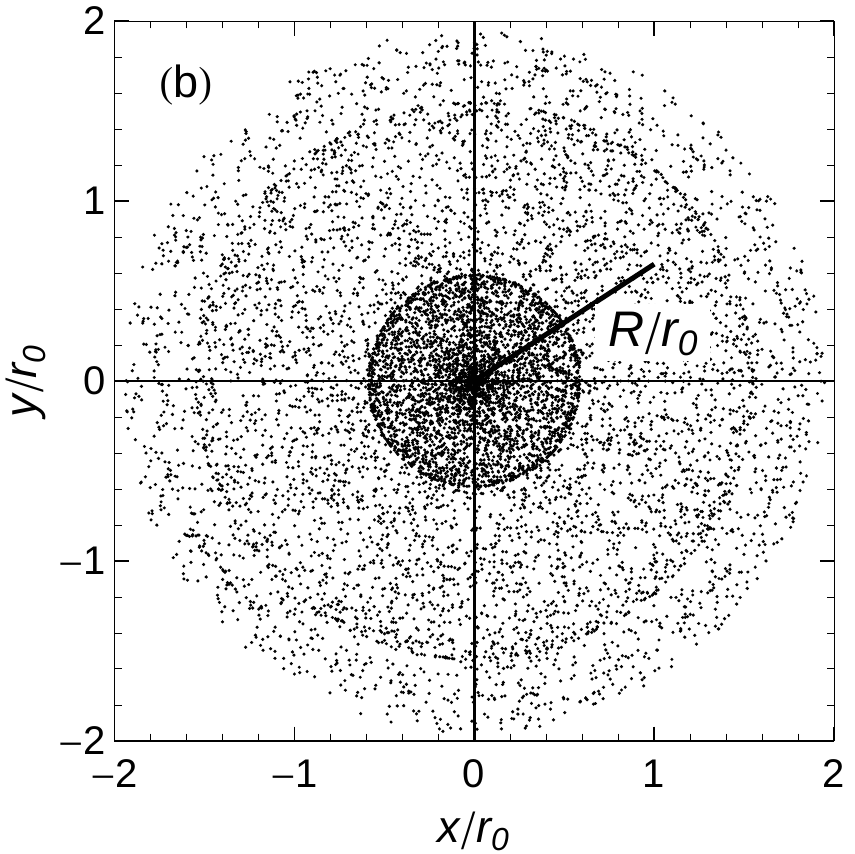}
\caption[] {(a) Impact loci curve at the detector plane for polar
emission angle variations in the interval $1 \geq\cos\theta\geq
0.2$ with decrements $\Delta\cos\theta = 0.001$. It is called in
the text trajectory. Azimuthal emission angle $\varphi$ = 0. (b)
Scatter plot for $\cos\theta$ and $\varphi$ randomly distributed.
Parameters of Ref. \cite{SjuB15} taken: $B$ = 0.326 Tesla, $pc$ =
0.976 MeV, $r_0 = p/(e B)$ = 0.01 m, $z_0$ = 0.10 m, and $z_0/r_0$
= 10.} \label{Trajectory}
\end{figure}
%%%%%%%%%%%%%%%%%%%%%%%%%%%%%%%%%%%%%%%%%%%%%%%%%%%%%%%%%%%%%%%%

\section{Radial distribution at the detector plane}
\label{MathBackbround} For the sake of convenience the
nomenclature of Sjue et al., \cite{SjuB15} will be adapted in the
following. A right-handed coordinate system is defined with the
magnetic field $\vec{B}$ coinciding with the $\hat{z}$ direction.
The detector is placed in the (x,y) plane at a distance $z_0$ from
the origin of the coordinate system in which the point source is
located. The charged particle starts with polar angle $\theta$ and
azimuthal angle $\varphi$, the latter defined with respect to the
$y$ axis. The point of impact at the detector plane is given by
\cite[Eq. (4)]{SjuB15}
\begin{eqnarray}
\frac{\vec r}{r_0} = \sqrt {1 - {{\cos }^2}\theta } \left\{ {\hat
x\left[ {\left( {1 - \cos \Big(\frac{z_0}{{{r_0}\cos \theta
}}\Big)} \right)\cos \varphi +
\sin \Big(\frac{z_0}{{{r_0}\cos \theta }}\Big)\sin \varphi } \right] + } \right.\nonumber\\
+ \left. {\hat y\left[ { - \left( {1 - \cos
\Big(\frac{z_0}{{{r_0}\cos \theta }}\Big)} \right)\sin \varphi +
\sin \Big(\frac{z_0}{{{r_0}\cos \theta }}\Big)\cos \varphi }
\right]} \right\} \label{Radiusvector}
\end{eqnarray}
with $r_0 = p/(q \cdot B)$ the maximum projected orbital radius,
$p$ the momentum, and $q$ the charge of the particle. Fig.
\ref{Trajectory} (a) depicts the loci the radius vector $\vec
r/r_0$ traverses according to Eq. (\ref{Radiusvector}) if the
cosine of the polar emission angle $\theta$ is decremented in
steps $\Delta\cos\theta = 0.001$. This impact loci curve must be
distinguished from the projected orbit of the electron on its way
from the target to the detector and will be called in the
following trajectory.

The radius coordinate in the (x,y) plane reads \cite[Eq.
(12)]{SjuB15}
\begin{eqnarray}
\frac{R(\cos \theta)}{r_0} =\sqrt{\left(\frac{x}{r_0}\right)^2 +
\left(\frac{y}{r_0}\right)^2} = \sqrt { 2\big(1 - {{\cos
}^2}\theta\big)\bigg(1 - \cos \Big(\frac{{{z_0}}}{{{r_0}\cos
\theta }}\Big)\bigg)}. \label{REq}
\end{eqnarray}
The differential probability $dP_c$ per normalized differential
radius interval $d(R/r_0)$ is
\begin{eqnarray}
\frac{dP_c(\cos\theta)}{dR/r_0}=\frac{dP_c(\cos\theta)}{d\cos\theta}\cdot\frac{d\cos\theta}{d(R/r_0)}=
\frac{dP_c/d\cos\theta}{d(R/r_0)/d\cos\theta}.
\end{eqnarray}
For the example of an isotropically emitting source with $dP(\cos
\theta)/d\cos\theta$ = 1 and $d(R/r_0)/d\cos\theta$ calculated
from Eq. (\ref{REq}) one obtains after some algebraic
manipulations
\begin{eqnarray}
\frac{{dP_c(\cos \theta)}}{{dR/{r_0}}} = \frac{{{r_0}{{\cos
}^2}\theta \sqrt {1 - {{\cos }^2}\theta } }}{{\left| {\left( {1 -
{{\cos }^2}\theta } \right){z_0}\cos
\big(\frac{{{z_0}}}{{2{r_0}\cos \theta }}\big) + 2{r_0}{{\cos
}^3}\theta \;\sin \big(\frac{z}{{2{r_0}\cos \theta }}\big)}
\right|}}, \label{dPdREq}
\end{eqnarray}
or normalized to the unit area
\begin{eqnarray}
\frac{{dP_c(\cos \theta)}}{dA}=\frac{1}{2\pi R
r_0}\frac{{dP_c(\cos \theta)}}{{dR/{r_0}}}. \label{dPdAEq}
\end{eqnarray}
It can be shown that Eq. (\ref{dPdAEq}) agrees with Eq. (14) of
Ref. \cite[]{Dub15}.

\begin{figure}[tb]
%\hspace{3.0 cm}
\centering
    \includegraphics[angle=0,scale=0.65,clip] {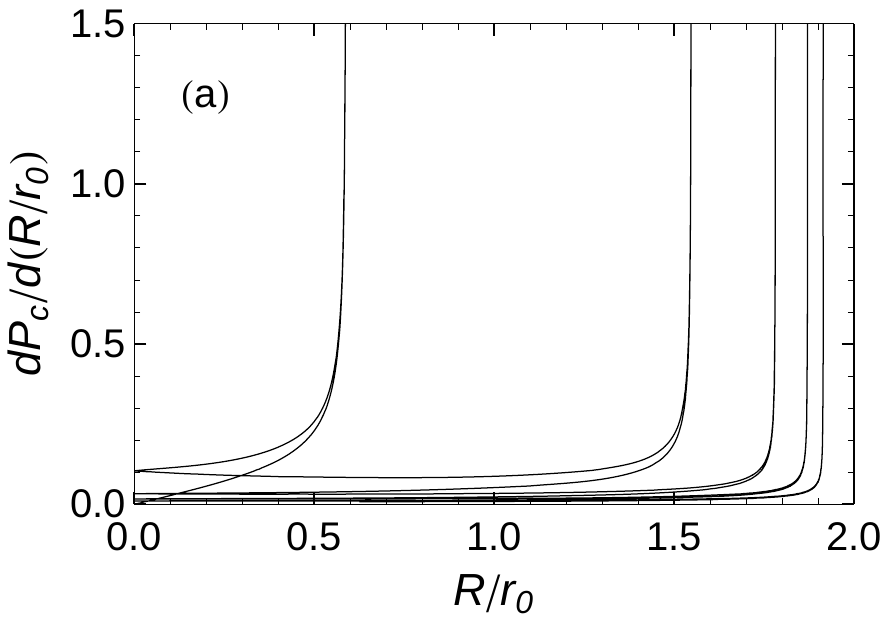}
    \includegraphics[angle=0,scale=0.65,clip] {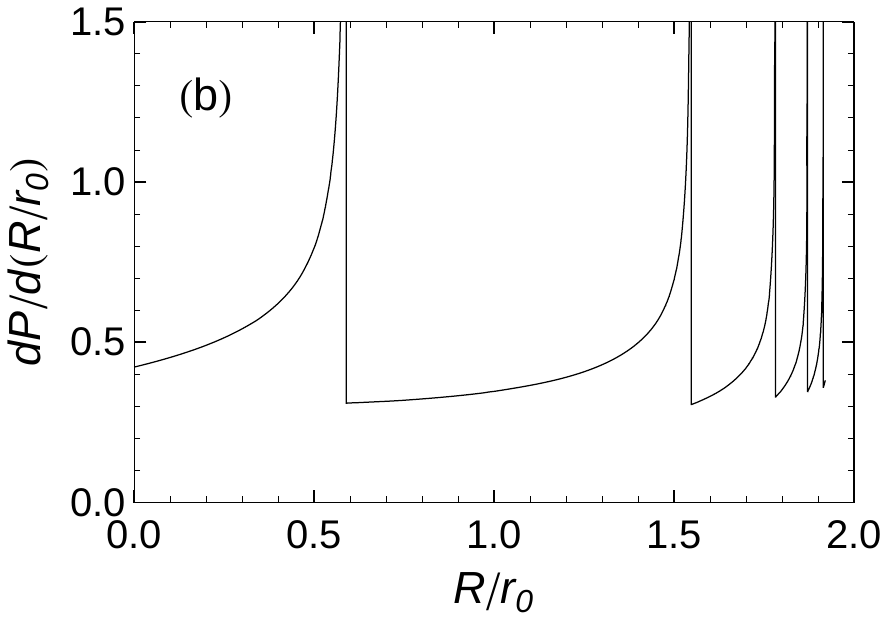}
\caption[] {(a) Parametric representation of the differential
probability $dP_c/(dR/r_0)$ as function of the normalized radius
coordinate $R/r_0$. The parameter $\cos \theta$ has been varied in
the interval \{1, 0.27\}, i.e. 73 \% of the emitted intensity has
been exhausted. For $\cos \theta$ = 1 the curve starts at the
origin, evolves with decreasing $\cos \theta$ to the first spike
which corresponds to the largest radius of the inner trajectory in
Fig. \ref{Trajectory} (a), returns to $R/r_0$ = 0 and evolves from
there to the next spike. (b) Sum of 500 trajectories as function
of the the normalized radius coordinate $R/r_0$. Magnetic field
and geometrical parameters the same as quoted in Fig.
\ref{Trajectory}.} \label{dPdRFig}
\end{figure}

Treating $\cos{\theta}$ as a free parameter, both the radial
detector coordinate $R(\cos{\theta})$ \emph{and}
$dP_c(\cos{\theta})/dR$ can be calculated with Eq. (\ref{REq}) and
(\ref{dPdREq}), respectively. If the parameter $\cos{\theta}$ is
varied within the interval \{1,0\} one gets an impression how
$dP_c/d(R/r_0)$ evolves as function of $R/r_0$. A corresponding
parametric plot is depicted in Fig. \ref{dPdRFig} (a). Shown are
branches which start at $R/r_0$ = 0 and end again at $R/r_0$ = 0
for each closed trajectory at the detector plane depicted in Fig.
\ref{Trajectory} (a). However, what is wanted, is the sum of all
individual contribution for a chosen normalized radius $R/r_0$.
This is the sum
\begin{eqnarray}
\frac{{dP\left( R/r_0 \right)}}{{dR/{r_0}}} = \sum\limits_{n =
{m}}^\infty {\left( {\frac{{d{P_{\mathop{\rm c}\nolimits} }\left(
{\left. {\cos \theta } \right|_n^ > } \right)}}{{dR/{r_0}}} +
\frac{{d{P_{\mathop{\rm c}\nolimits} }\left( {\left. {\cos \theta
} \right|_n^ < } \right)}}{{dR/{r_0}}}} \right)} \label{dPdRSumEq}
\end{eqnarray}
with $\left.\cos \theta \right|_n^>$ and $\left.\cos \theta
\right|_n^<$ the two solutions of the equation
\begin{eqnarray}
{\sqrt {2\big( {1 -{{\cos }^2}\theta } \big)\left( {1 - \cos
\Big(\frac{{{z_0}}}{{{r_0}\cos \theta }}\Big)} \right)}  =
\frac{R}{{{r_0}}}} \label{rootEq}
\end{eqnarray}
for the n$^{th}$ trajectory in the interval
\begin{eqnarray}
\frac{{{z_0}}}{{{r_0}2\pi (n+n_f)}} < \cos \theta \le \min
\Big[\frac{{{z_0}}}{{{r_0}2\pi (n - 1 + n_f)}},1 \Big],
\label{inequalEq}
\end{eqnarray}
and $n_f=\mbox{floor}(z_0/(r_0 2 \pi))$. The lower limit of the
summation $m$ is the smallest integer for which $R/R_n > 1$ holds
with $R_n$ the maximum radius of the n$^{th}$ trajectory in the
interval defined by the inequality (\ref{inequalEq}).

%%%%%%%%%%%%%%%%%%%%%%%%%%%%%%%%%%%%%%%%%%%%%%%%%%%%%%%%%%%%%%%%
%%%%%%%%%%%%%%%%%%%%%%%%%%%%%%%%%%%%%%%%%%%%%%%%%%%%%%%%%%%%%%%%

\section{Results and Discussion}\label{ResultsDiscussion}
In Fig.\ref{dPdRFig} (b) numerical results on the basis of Eq.
(\ref{dPdRSumEq}) are depicted. Totally 500 trajectories at the
detector plane of Fig. \ref{Trajectory} (a) have been taken into
account. The accuracy of the calculation at $R/r_0$ = 1.92 and
$R/r_0 = 0.6$ has been estimated to be in the order of 1.8 \%, and
0.6 \%, respectively. The accuracy can be improved by extending
the summation in Eq. (\ref{dPdRSumEq}) over more trajectories.
However, it should be mentioned that the series expansion Eq.
(\ref{dPdRSumEq}) converges rather slowly.

%%%%%%%%%%%%%%%%%%%%%%%%%%%%%%%%%%%%%%%%%%%%%%%%%%%%%%%%%%%%%%%%
\begin{figure}[tb]
%\hspace{3.0 cm}
\centering
    \includegraphics[angle=0,scale=0.65,clip] {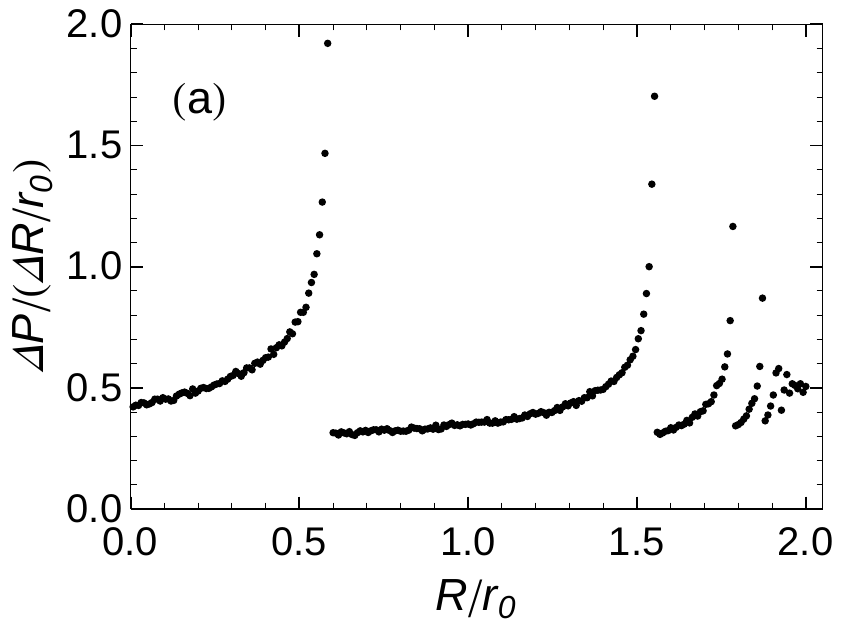}
    \includegraphics[angle=0,scale=0.65,clip] {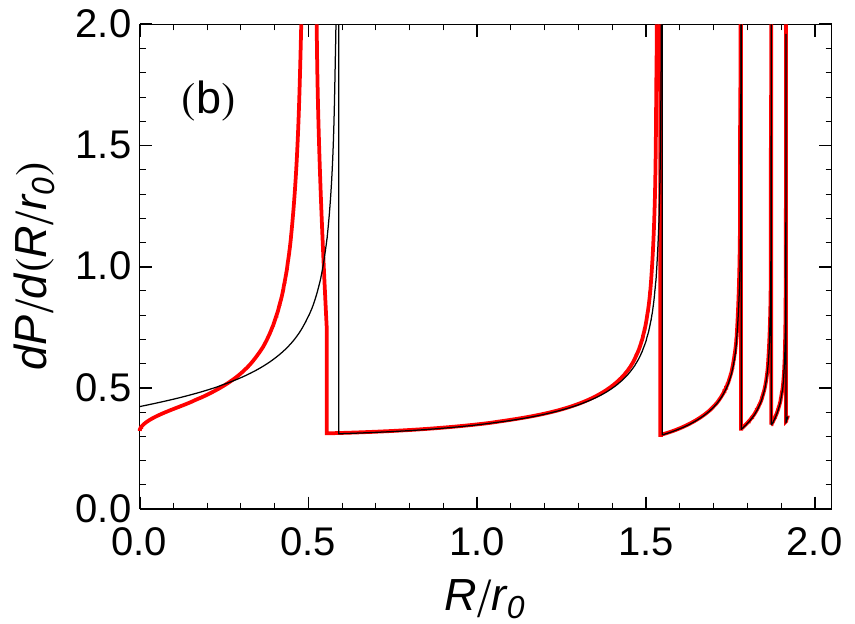}
\caption[] {(a) Random distribution as shown in Fig.
\ref{Trajectory} (b) analyzed at the detector plane with a radial
resolution of 0.8 \% or for the treated example of 80~$\mu$m. (b)
Comparison of the numerical true PSF of Dubbers \cite{Dub15} with
the approximations presented in chapter 3 in red with the exact
solution in black.} \label{dPdRCompareDub}
\end{figure}
%%%%%%%%%%%%%%%%%%%%%%%%%%%%%%%%%%%%%%%%%%%%%%%%%%%%%%%%%%%%%%%%

In the interval $1.92<R/r_0\leq 2$ infinitely many spikes appear.
It is not necessary to treat them in a mathematically exact
manner. This fact is demonstrated in Fig. \ref{dPdRCompareDub}
(a). A number of 10$^6$ impact points at the detector plane were
generated by randomly distributing $\cos\theta$ and $\varphi$ in
Eq. (\ref{Radiusvector}). A small sample is shown in Fig.
\ref{Trajectory} (b). The generated distribution has been analyzed
with an virtual detector of 0.8 \% spatial resolution in radial
direction, corresponding for the chosen example with $r_0$ = 1 cm
to 80~$\mu$m. It can clearly be seen that all the spikes in the
mentioned interval result in a mean converging against 0.5. It
should be mentioned that the same argument holds more or less for
a finite beam spot size of the same order of magnitude. Fig.
\ref{dPdRCompareDub} (a) is fully in accord with Fig. 4 of Sjue et
al. \cite{SjuB15}. Probably the procedure applied by the authors
to generate their Fig. 4 was nothing else than what has been just
described here.

In Fig. \ref{dPdRCompareDub} (b) the exact results are compared
with the approximations elaborated by Dubbers \cite{Dub15}. In due
distance from the spikes and for outer trajectories in Fig.
\ref{Trajectory} (a) with many revolutions, his approximation
apparently seems to be rather good. For the important inner
trajectory the approximation is rather poor, and even
normalization is not preserved.

Finally the question should be addressed whether the
considerations on the PSF of Ref. \cite{SjuB15, Dub15} can be
applied for field configurations other than homogeneous ones. It
has been pointed out by Dubbers \cite[ch. 5]{Dub15} that in
axially symmetric, continuously descending magnetic fields the
general formulas remain valid after replacement of $R$ by
$R\cdot\sqrt{B(z_0)/B(0)}$ with $B(0)$ and $B(z_0)$ the magnetic
fields on-axis at the source and detector position, respectively.
It is certainly true that the underlying adiabatic invariance
considerations makes some valuable statements on the movement of
charged particles in inhomogeneous magnetic fields \cite[p.
592f]{Jac99}, see also Ref. \cite{KotB70}. However, whether in the
fundamental Eq. (\ref{REq}) the phase $z_0/(r_0\cos\theta)$ of the
cosine function can simply be replaced by $\sqrt{B(0)
B(z_0)}~z_0/[(B\rho)_p\cos\theta]$, with $(B\rho)_p = p/q$,
maintaining otherwise the functional dependence is questionable
and needs to be discussed in terms of quantitative accuracy
considerations. For more general magnetic field configurations it
seems unlikely to find analytical solutions equivalent to Eq.
(\ref{REq}) and (\ref{dPdREq}) and experimentalists would be well
advised to investigate their instruments from the beginning by
Monte-Carlo simulations performed with exact orbit calculations.

\section{Conclusions} \label{Conclusions}
A method has been described with which mono-energetic point spread
functions can be calculated with arbitrary accuracy for a
homogeneous magnetic guiding field. It has been shown by
Monte-Carlo simulations that a finite detector resolution or a
finite target spot size smear out the singularities for
trajectories in the detector plane originating from polar emission
angles approaching $\theta = \pi/2$. The results of Sjue et al.
\cite{SjuB15} are fully in accord with the results obtained in
this paper. Although Dubbers \cite{Dub15} presents the correct
parameter representation for the probability density function,
which is not explicitly quoted by Sjue et al. \cite{SjuB15}, his
analytical approximations for the singularities appear for the
innermost trajectories to be rather inaccurate with even a
significant violation of the normalization.

In any case, the subject addressed in Ref. \cite{SjuB15, Dub15} is
appealing and certainly beneficial for intuitional and educational
purposes.

\section*{Acknowledgements} Calculations have been performed with
the Wolfram Mathematica8.0 package. Pictures were prepared with
the LevelScheme scientific figure preparation system by M. A.
Caprio, Department of Physics, University of Notre Dame, Version
3.53 (January 10, 2013) \cite{Cap05}.

\section*{References}

\bibliography{bibfilePSF}

\end{document}